\documentclass[a4paper,11pt]{article}

\usepackage{jcappub} 

\usepackage[T1]{fontenc}

\title{Quintessence versus phantom dark energy: the arbitrating power of current and future observations}

\author[a]{B. Novosyadlyj,}
\author[a]{O. Sergijenko,}
\author[b]{R. Durrer}
\author[c]{and V. Pelykh}

\affiliation[a]{Astronomical Observatory of Ivan Franko National University of Lviv,\\ Kyryla i Methodia str., 8, Lviv, 79005, Ukraine}
\affiliation[b]{Universit\'e de Gen\`eve, D\'epartement de Physique Th\'eorique and CAP,\\ 24 quai Ernest-Ansermet, CH-1211 Gen\`eve 4, Switzerland}
\affiliation[c]{Ya. S. Pidstryhach Institute for Applied Problems of Mechanics and Mathematics,\\ Naukova str., 3-b, Lviv, 79060, Ukraine}

\emailAdd{novos@astro.franko.lviv.ua}
\emailAdd{olka@astro.franko.lviv.ua}
\emailAdd{ruth.durrer@unige.ch}
\emailAdd{pelykh@iapmm.lviv.ua}

\abstract{We analyze the possibility to distinguish between quintessence and phantom scalar field models of dark energy using
observations of luminosity distance moduli of SNe Ia, CMB anisotropies and polarization, matter density perturbations and
baryon acoustic oscillations. Among the present observations only Planck data on CMB anisotropy and SDSS DR9 data on baryon acoustic 
oscillations may be able to decide between quintessence or phantom scalar field  models, however for each model a set of best-fit 
parameters exists, which matches all data with similar goodness of fit. We compare the relative differences of best-fit model 
predictions with observational uncertainties for each type of data and we show that the accuracy of SNe Ia luminosity distance 
data is far from the one necessary to distinguish these types of dark energy models, while the CMB data (WMAP, ACT, SPT and especially 
Planck) are close to being able to reliably distinguish them. Also an improvement of the large-scale structure data (future releses of 
SDSS BOSS and e.g. Euclid or BigBOSS) will enable us to surely decide between quintessence and phantom dark energy.}

\begin{document}
\maketitle
\flushbottom

\section{Introduction}

Scalar field models of dark energy are among the most promising and best elaborated ones to match observations of the accelerated expansion of the Universe. This is reflected in the numerous papers, reviews and text books on the subject
(see for example \cite{Copeland2006,Turner2007,GRG2008,Frieman2008,Lukash2008,Caldwell2009,Cai2010,Blanchard2010,Sapone2010,Amendola2010,Wolchin2010} and citing therein). However, despite  good agreement of theoretical predictions with modern observational data a number of problems remain unresolved and the prospects for their solution are unclear. Among them is  the question
whether dark energy is quintessence, phantom, quintom, k-essence or of some other form. For each of these types of dark energy, cosmological best-fit parameters exist, for which the differences of maximum likelihoods are statistically insignificant for the same datasets. The identification of dark energy type is important in order to establish the properties of dark energy in our Universe, its physical nature, its origin and its couplings to other fundamental fields and particles. Without this knowledge, e.g., we cannot predict the future of our Universe.

For example, in the paper \cite{Novosyadlyj2011,Sergijenko2011} we have shown that quintessence models of dark energy  with decreasing and increasing equation of state (EoS) parameter (freezing or thawing quintessence) cannot be distinguished by current observations, but could be so by future data. In the recent papers \cite{Novosyadlyj2012,Wang2012} it has been shown also that some datasets slightly prefer phantom models over quintessence, while in contrary other datasets prefer quintessence. However, in all the cases considered, the differences in the maximum likelihoods are not statistically significant. 

The goal of this paper is to estimate the possibility to distinguish between quintessence and phantom scalar field models using the current and 
future observational data.  

\section{Scalar field dark energy and best-fit cosmological models}

We suppose that our Universe is spatially flat and filled with the non-relativistic particles (cold dark matter and baryons), 
relativistic particles (the thermal cosmic microwave background (CMB) and massless neutrino) and a minimally coupled scalar field with given Lagrangian. The scalar field is specified as follows: (i) its Lagrangian is canonical, $L_{de}=\pm X-U(\phi)$ ($X=\dot\phi^2/2$ is the  kinetic term and $U(\phi)$ is the potential), with ``+`` for a quintessence scalar field (QSF) and ''-`` for a phantom scalar field (PSF). The effective sound speed $c_{s\,(de)}^2\equiv \delta p_{de}/\delta\rho_{de}$ (speed of propagation of the scalar field perturbations) is equal to the speed of light $c$ for both cases; (ii) its equation of state is $p_{de}=w_{de}c^2\rho_{de}$ and the time derivative of the dark energy pressure is proportional to the time derivative of 
its energy density, $\dot{p}_{de}=c_a^2c^2\dot{\rho}_{de}$.  The coefficient $c_a^2$, often called ``squared adiabatic sound speed''\footnote{It corresponds only formally to the adiabatic sound speed in thermodynamics.}, is constant. Integrating this equation we obtain the generalized linear barotropic equation of state $p_{de}=c_a^2\rho_{de}+C$, where $C$ is a constant, that is why we call such scalar field  barotropic.

We solve Einstein's equations for the background dynamics and the Einstein-Boltzmann system of linear perturbation equations in the synchronous gauge for the evolution of the perturbations. The background Universe is assumed to be the spatially flat homogeneous and isotropic with Friedmann-Robertson-Walker (FRW) metric, $ds^2=g_{ij} dx^i dx^j =a^2(\eta)(d\eta^2-\delta_{\alpha\beta} dx^{\alpha}dx^{\beta}),$ 
where $\eta$ is conformal time defined by $cdt=a(\eta)d\eta$ and $a(\eta)$ is the scale factor, normalized to 1 today (below we set $c=1$). We assume that dark energy and dark matter do not interact. The energy-momentum conservation for dark energy then determines $w_{de}$ and $\rho_{de}$ as functions of the scale factor $a$,
\begin{equation}
 w_{de}=\frac{(1+c_a^2)(1+w_0)}{1+w_0-(w_0-c_a^2)a^{3(1+c_a^2)}}-1, \quad
\rho_{de}=\rho_{de}^{(0)}\frac{(1+w_0)a^{-3(1+c_a^2)}+c_a^2-w_0}{1+c_a^2} \,.\label{w_rho}
\end{equation}
Using the Friedmann equations we obtain the Hubble and deceleration parameters
\begin{equation}
 H=H_0\sqrt{\Omega_r a^{-4}+\Omega_m a^{-3}+\Omega_{de}f(a)}, \quad
 q=\frac{1}{2}\frac{2\Omega_r a^{-4}+\Omega_m a^{-3}+(1+3w_{de})\Omega_{de}f(a)}
{\Omega_r a^{-4}+\Omega_m a^{-3}+\Omega_{de}f(a)},\label{H_q}
\end{equation}
where $f(a)=\rho_{de}/\rho_{de}^{(0)}$, $\rho_{de}^{(0)}$ is current value of dark energy density, $w_0$ is EoS parameter $w_{de}$ today, $H_0$ is current value of Hubble parameter (Hubble constant) and $\Omega_r\equiv \rho_{r}^{(0)}/\rho_{tot}^{(0)}$, $\Omega_m\equiv \rho_{m}^{(0)}/\rho_{tot}^{(0)}$, 
$\Omega_{de}\equiv \rho_{de}^{(0)}/\rho_{tot}^{(0)}$ are the dimensionless density parameters of the  relativistic, non-relativistic and dark energy components correspondingly. The first Friedmann equation requires $\Omega_r +\Omega_m+ \Omega_{de}=1$. The matter density parameter is the sum of cold dark matter and baryons, $\Omega_m\equiv\Omega_{cdm}+\Omega_b$, and $\rho_{tot}^{(0)}\equiv \rho_{r}^{(0)}+\rho_{m}^{(0)}+\rho_{de}^{(0)}$. 

Since the light neutrinos have become non-relativistic by today, the density parameter of the present relativistic component is given only by the photon density which is accurately determined by the value of the CMB temperature, $\Omega_r=\Omega_{\gamma}=16\pi Ga_{SB}T_0^4/3H_0^2=2.49\cdot10^{-5}h^{-2}$ and can be neglected in the matter and dark energy dominated epochs. The density parameters of the other components are less well known and depend somewhat on the model of dark energy\footnote{Here and below $h\equiv H_0/100\textrm{km/s\,Mpc}$}. The scalar field affects the expansion of the Universe, it causes accelerated expansion when $|(1+3w_{de})\Omega_{de}f(a)|>\Omega_m a^{-3}$ as follows from eqs.~(\ref{H_q}). 

Using expressions (\ref{w_rho}) and (\ref{H_q}) we can compute the ``luminosity distance - redshift`` or ``angular diameter distance - redshift'' relations to determine all the above-mentioned parameters by comparison with corresponding observational data on standard candles (supernovae type Ia, $\gamma$-ray bursts or other) and standard rulers (positions of the CMB acoustic peaks, baryon acoustic oscillations, X-ray gas in clusters or other).  

We assume the standard paradigm of large scale structure formation: (i) it is formed from stochastic, adiabatic, Gaussian scalar perturbations generated in the early Universe, (ii) the initial power spectrum of radiation and matter density perturbations is power-law, $P_i(k)=A_sk^{n_s}$, where $A_s$ and $n_s$ are the amplitude and spectral index ($k$ is wave number). The scalar field cannot be perfectly smooth, it is perturbed by gravitational influence of matter-radiation inhomogeneities or has its own initial fluctuations, generated in the early Universe. 

The system of linear differential equations for the evolution of quintessence and phantom scalar field perturbations and their numerical solutions are analyzed in our previous papers \cite{Novosyadlyj2012,Novosyadlyj2009,Novosyadlyj2010}. The main conclusions are as follows: (i) the amplitude of scalar field density perturbations at any epoch depends strongly on parameters of barotropic scalar field $\Omega_{de}$, $w_0$, $c_a^2$ and $c_s^2$; (ii) although the density perturbations of dark energy at the current epoch are significantly smaller than matter density perturbations, they leave noticeable imprints in the matter power spectrum, which can be used to constrain the scalar field parameters. 

The linear  power spectrum of each component  can be computed by numerical integration of the Einstein-Boltzmann equations \cite{Ma1995,Durrer2001,Novosyadlyj2007,Durrer2008} using publicly available codes such as CMBFAST \cite{cmbfast96,cmbfast99}, CMBEasy \cite{Doran2005}, CAMB \cite{camb,camb_source} or CLASS \cite{Lesgourgues2011,Blass2011,class_source}, 
$$P_{lin}(k)=P_i(k)T^2(k;\Omega_r,\Omega_b,\Omega_{cdm},\Omega_{de})\,. $$
 $T(k)$ is the transfer function, for the cosmological model with the given parameters.  In all the computations presented here and in recent papers, we use CAMB with corresponding modifications for the quintessence/phantom barotropic scalar field as dark energy. 
 Comparison of computed matter density power spectra with the ones obtained from galaxy surveys constrains the parameters of the models discussed here.

The same Einstein-Boltzmann equations and codes  also determine the angular power spectra of CMB temperature anisotropies, $C^{TT}_{\ell}$, and polarization, $C^{EE}_{\ell}$, as well as their correlations,  $C^{TT}_{\ell}$, which can be compared with WMAP data to constrain  the cosmological and DE parameters mentioned above. The calculation of CMB anisotropies and polarization  requires also the knowledge of the reionization history of the Universe, which depends on complicated non-linear effects of structure formation. This is parameterized by the value of optical depth from current epoch to decoupling (at redshift $z_{dec}$) caused by Thomson scattering. It is denoted by $\tau_{dec}$ and is also fitted by the data. 
As it is supported by numerous papers (see for example reviews \cite{Copeland2006,Turner2007,GRG2008,Frieman2008,Lukash2008,Caldwell2009,Cai2010,Blanchard2010,Sapone2010,Amendola2010,Wolchin2010}
and citing therein), the CMB data are the most important ones in the determination of cosmological parameters.  
The parameter estimation from the data is performed using  publicly available Markov chain Monte Carlo (MCMC) codes~\cite{cosmomc,cosmomc_source}. 

\begin{table}[tbp]
\centering
  \caption{The best-fit values, mean values and 2$\sigma$ marginalized limits of parameters of cosmological models
with QSF ($\mathbf{q}$) and PSF ($\mathbf{p}$) as dark energy determined by the MCMC technique using 2 different observational datasets: WMAP7 {+} HST {+} BBN {+} BAO {+} SN SDSS SALT2 ($\mathbf{q}_1$, $\mathbf{p}_1$) and WMAP7 {+} HST {+} BBN {+} BAO {+} SN SDSS MLCS2k2 ($\mathbf{q}_2$, $\mathbf{p}_2$). We denote the rescaled energy density of the component $X$ by $\omega_X \equiv \Omega_Xh^2$.}
  \medskip\footnotesize{
  \begin{tabular}{|c|c|c|c|c|c|c|c|c|}
    \hline 
    &\multicolumn{2}{c|}{}&\multicolumn{2}{c|}{}&\multicolumn{2}{c|}{}&\multicolumn{2}{c|}{}\\
    Parameters&\multicolumn{2}{c|}{QSF+CDM}&\multicolumn{2}{c|}{PSF+CDM}&\multicolumn{2}{c|}{QSF+CDM}&\multicolumn{2}{c|}{PSF+CDM}\\
    &\multicolumn{2}{c|}{}&\multicolumn{2}{c|}{}&\multicolumn{2}{c|}{}&\multicolumn{2}{c|}{}\\
    \cline{2-9}
    &&&&&&&&\\
     &$\mathbf{q}_1$&2$\sigma$ c.l.&$\mathbf{p}_1$&2$\sigma$ c.l.&$\mathbf{q}_2$&2$\sigma$ c.l.&$\mathbf{p}_2$&2$\sigma$ c.l.\\
    &&&&&&&&\\
    \hline
    &&&&&&&&\\
    $\Omega_{de}$&0.730& 0.725$_{- 0.030}^{+ 0.028}$&0.723& 0.725$_{- 0.030}^{+ 0.027}$&0.702& 0.700$_{- 0.034}^{+ 0.031}$&0.692& 0.700$_{- 0.034}^{+ 0.032}$\\&&&&&&&&\\
    $w_0$&-0.996&-0.966$_{- 0.034}^{+ 0.060}$& -1.043&-1.077$_{- 0.108}^{+ 0.077}$& -0.83&-0.873$_{- 0.127}^{+ 0.126}$&-1.002&-1.034$_{- 0.063}^{+ 0.034}$\\&&&&&&&&\\
    $c_a^2$& -0.022&-0.591$_{- 0.409}^{+ 0.591}$& -1.12&-1.349$_{- 0.226}^{+ 0.349}$& -0.88&-0.713$_{- 0.287}^{+ 0.423}$&-1.19&-1.318$_{- 0.239}^{+ 0.318}$\\&&&&&&&&\\
    10$\omega_b$& 0.226& 0.226$_{- 0.010}^{+ 0.011}$&0.223& 0.224$_{- 0.010}^{+ 0.010}$&0.226& 0.227$_{- 0.011}^{+ 0.011}$&0.223& 0.223$_{- 0.010}^{+ 0.010}$\\&&&&&&&&\\
    $\omega_{cdm}$& 0.110& 0.110$_{- 0.008}^{+ 0.008}$&0.115& 0.115$_{- 0.007}^{+ 0.008}$&0.108& 0.110$_{- 0.010}^{+ 0.009}$&0.119& 0.119$_{- 0.007}^{+ 0.007}$\\&&&&&&&&\\
    $h$&0.702& 0.695$_{- 0.027}^{+ 0.027}$&0.704& 0.709$_{- 0.026}^{+ 0.027}$&0.663& 0.665$_{- 0.029}^{+ 0.030}$&0.678& 0.686$_{- 0.026}^{+ 0.026}$\\&&&&&&&&\\
    $n_s$& 0.974& 0.973$_{- 0.025}^{+ 0.027}$&0.965& 0.964$_{- 0.024}^{+ 0.024}$&0.971& 0.975$_{- 0.027}^{+ 0.027}$&0.965& 0.962$_{- 0.024}^{+ 0.023}$\\&&&&&&&&\\
    $\log(10^{10}A_s)$& 3.085& 3.081$_{- 0.067}^{+ 0.071}$&3.089& 3.091$_{- 0.063}^{+ 0.067}$&3.069& 3.082$_{- 0.069}^{+ 0.070}$&3.113& 3.098$_{- 0.063}^{+ 0.065}$\\&&&&&&&&\\
    $\tau_{rei}$&0.091& 0.090$_{- 0.024}^{+ 0.026}$&0.085& 0.086$_{- 0.022}^{+ 0.024}$&0.089& 0.090$_{- 0.024}^{+ 0.025}$&0.086& 0.084$_{- 0.022}^{+ 0.023}$\\&&&&&&&&\\
    \hline
    &\multicolumn{2}{c|}{}&\multicolumn{2}{c|}{}&\multicolumn{2}{c|}{}&\multicolumn{2}{c|}{}\\
    $-\log L$&\multicolumn{2}{c|}{3865.01}&\multicolumn{2}{c|}{3864.86}&\multicolumn{2}{c|}{3857.21}&\multicolumn{2}{c|}{3859.30}\\
    &\multicolumn{2}{c|}{}&\multicolumn{2}{c|}{}&\multicolumn{2}{c|}{}&\multicolumn{2}{c|}{}\\
    \hline
  \end{tabular}}
  \label{tab_bfp}
\end{table}
 
The cosmological model with quintessence/phantom barotropic scalar field with canonical Lagrangian has 9 free parameters $\Omega_b$, $\Omega_{cdm}$, $\Omega_{de}$, $w_0$, $c_a^2$, $H_0$, $A_s$, $n_s$, $\tau_{dec}$ in the general case and 8 when we restrict ourselves to spatially flat cosmological models, since $\Omega_{b}+\Omega_{cdm}+\Omega_{de}=1$.  To constrain these parameters we used the following datasets: 
\begin{enumerate}
\item CMB temperature fluctuations and polarization angular power spectra
from the 7-year WMAP observations (hereafter WMAP7) \cite{WMAP7a,WMAP7b,WMAP7c}; 
\item Baryon acoustic oscillations in the space distribution of galaxies from SDSS DR7 (hereafter BAO) \cite{Percival2010}; 
\item Hubble constant measurements from HST (hereafter HST) \cite{Riess2009};
\item Big Bang Nucleosynthesis prior on baryon abundance (hereafter BBN) \cite{Steigman2007,Wright2007};
\item Supernovae Ia luminosity distances from SDSS compilation (hereafter SN SDSS) \cite{Kessler2009}, determined using SALT2 method of 
light curve fitting \cite{Guy2007} (hereafter SN SDSS SALT2) and MLCS2k2 \cite{Jha2007} one (hereafter SN SDSS MLCS2k2). 
\end{enumerate}
 
In our previous paper \cite{Novosyadlyj2012} we have performed the Markov chain Monte Carlo (MCMC) analysis for two combined datasets: WMAP7 {+} HST {+} BBN {+} BAO {+} SN SDSS SALT2 (dataset 1) and WMAP7 {+} HST {+} BBN {+} BAO {+} SN SDSS MLCS2k2 (dataset 2)
to determine the best-fit values and confidence limits of the model parameters for QSF/PSF. The best-fit and mean values of 9 parameters as well as
their 95\% marginalized limits are presented in Table \ref{tab_bfp}. 
The constraints on $c_a^2$ from all data (except SN SDSS MLCS2k2 for quintessential fields) are weak \cite{Sergijenko2011,Novosyadlyj2012}. In the last row the minus logarithms of maximum of likelihood functions, $\chi^2=-log(L_{max})$, are shown for 
comparison. MCMC runs for models with QSF and PSF differ only by priors for $w_0$ and $c_a^2$: $-1< w_0\le 0$, $-1< c_a^2\le 0$ and 
$-2\le w_0<-1$, $-2\le c_a^2<-1$ correspondingly. The PSF+CDM model with best-fit parameters $\textbf{p}_1$ 
determined on the base of dataset 1 has a slightly lower $\chi^2$ than the QSF+CDM model with best-fit parameters $\textbf{q}_1$. 
In the case of dataset 2 the result is opposite: the QSF+CDM model with best-fit parameters $\textbf{q}_2$ matches the dataset 2 better than 
the PSF+CDM model with best-fit parameters $\textbf{p}_2$. 

In Ref.~\cite{Novosyadlyj2012} we have also determined the best-fit parameters of $\Lambda$CDM from the datasets 1 and  2 and we have found 
that in both cases the $\chi^2$ for $\Lambda$CDM ($\chi^2_{(1)}=3864.96$, $\chi^2_{(2)}=3859.15$) is between the corresponding values for QSF+CDM and PSF+CDM. This trend suggests that the 
best-fit parameters of QSF+CDM and PSF+CDM  (except for $c_a^2$) are determined reliably. But in both cases the differences of the $\chi^2$'s 
are not statistically significant. 
Hence, we have a degeneracy between the QSF and PSF models of dark energy and we shall elucidate here how strong this degeneracy is and whether 
the accuracy of  current or forthcoming data is sufficient for lift it. Observational constraints for another class of scalar fields with barotropic equation of state without peculiarities in the past are discussed in Appendix~\ref{app}.

The results for the cosmological parameters, especially $H_0$, $\Omega_{de}$, $w_{de}$ and $c_a^2$, presented in Table \ref{tab_bfp}, also indicate a tension between the two 
fitters SALT2 and MLCS2k2 applied to the same SNe Ia. This has already been highlighted and analyzed in Refs.~\cite{Kessler2009,Bengochea2011}, but up to now we have no decisive arguments for favor of one of the two lightcurve fitters.

In Ref.~\cite{Novosyadlyj2012} we have used also other sets of current observational data including the SNe Ia luminosity distances 
from SNLS3 \cite{snls3} and Union2.1 compilation \cite{union}, the BAO data from the WiggleZ Dark Energy Survey 
\cite{wigglez} and have shown the same low distinguishability of quintessence and phantom scalar field models of dark energy.
 
\section{Distinguishing quintessence and phantom field models of dark energy}\label{disting_sf}

\begin{figure}[tbp]
  \centering
  \includegraphics[width=.49\textwidth]{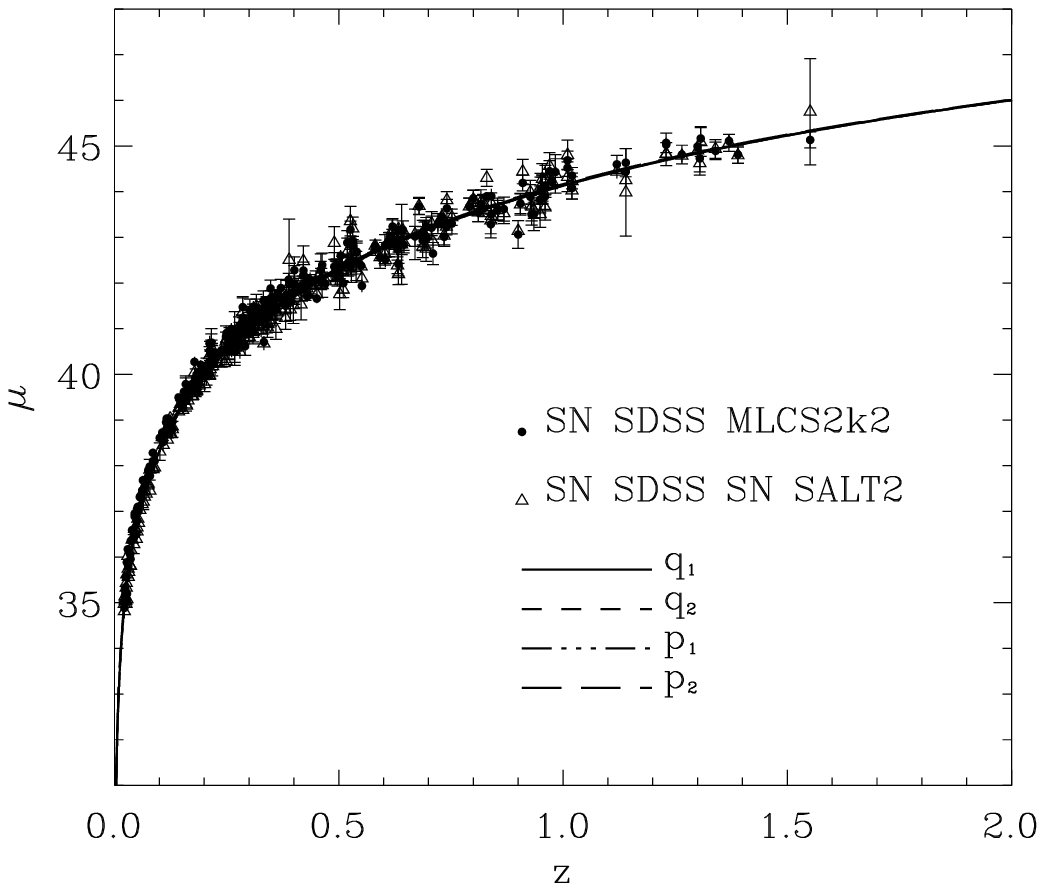}
  \includegraphics[width=.49\textwidth]{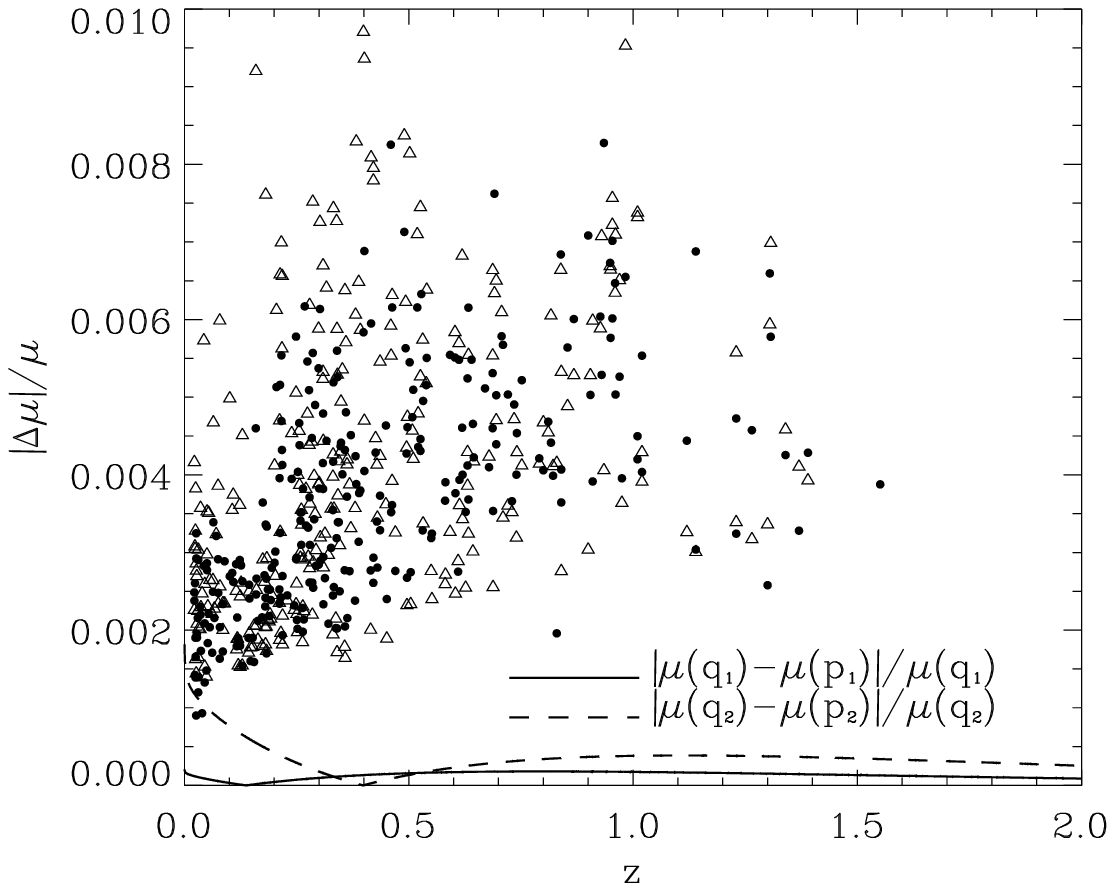}
  \includegraphics[width=.49\textwidth]{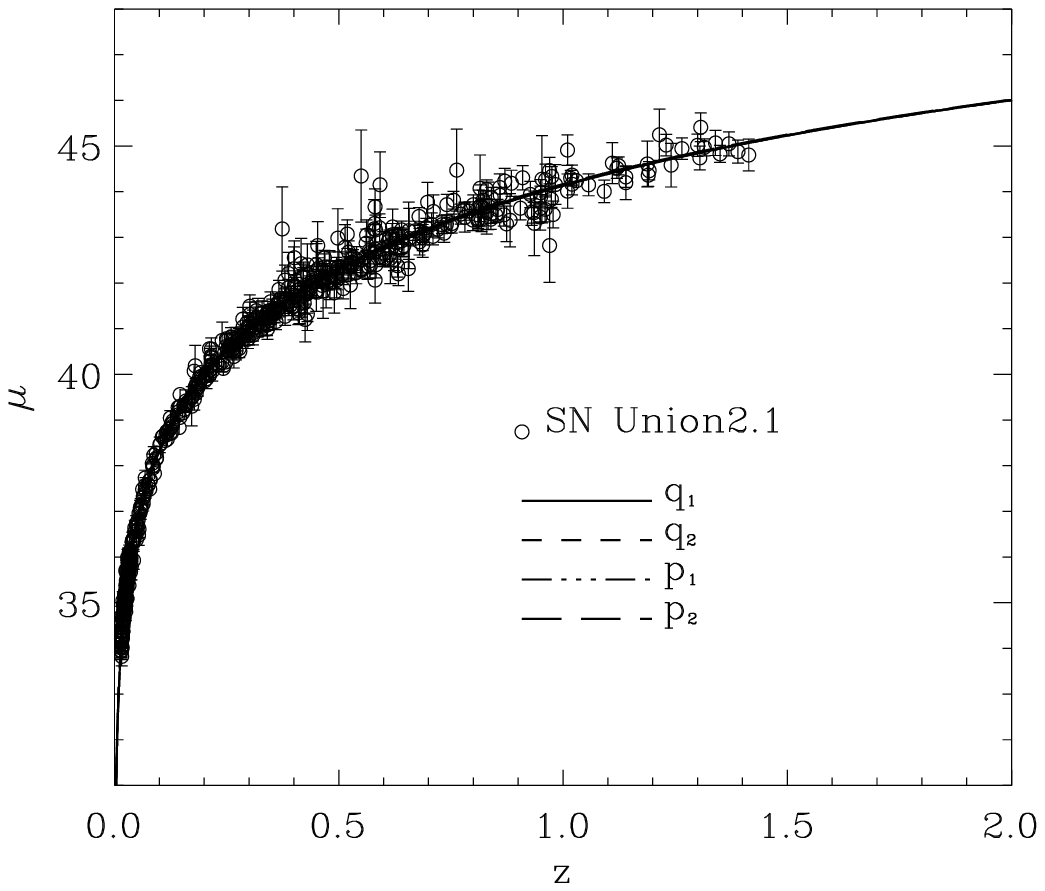}
  \includegraphics[width=.49\textwidth]{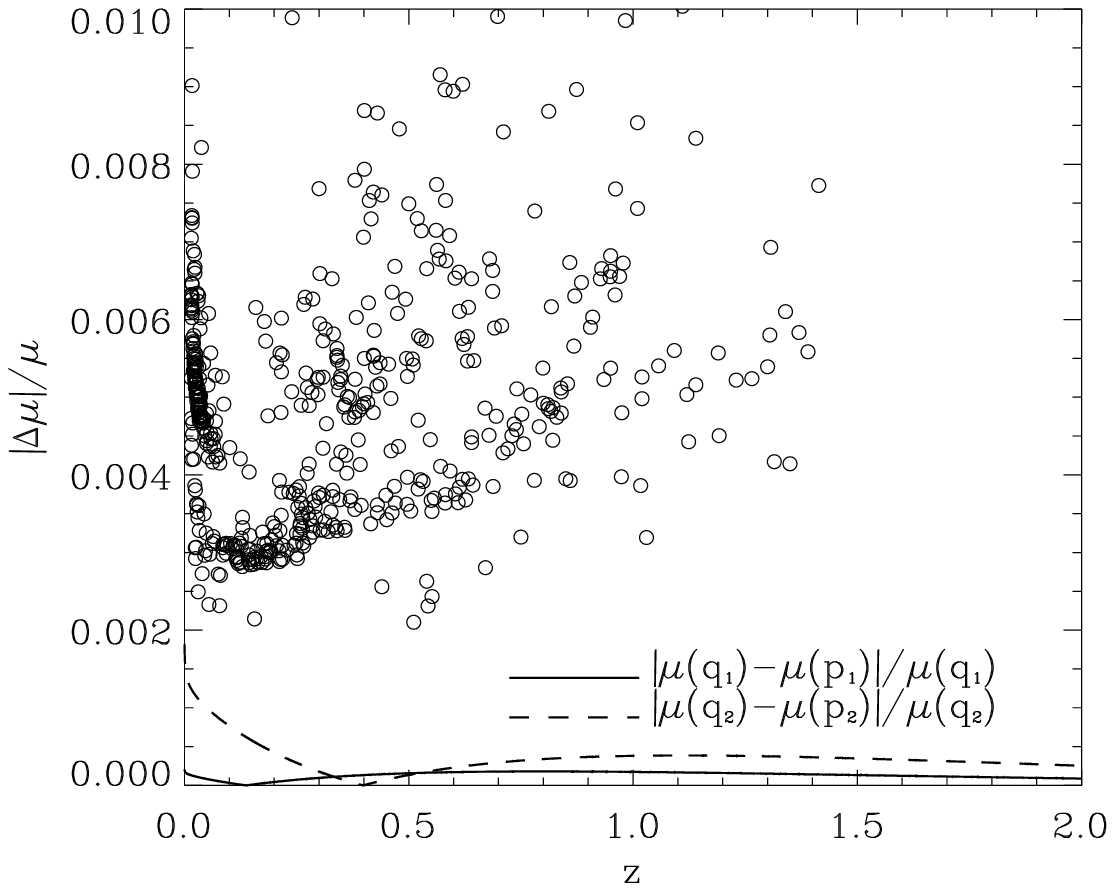}
  \caption{Left panel: the distance modulus $\mu\equiv m-M$ as function of the redshift $z$ in the models with best-fit 
parameters $\mathbf{q}_1$, $\mathbf{q}_2$, $\mathbf{p}_1$ and $\mathbf{p}_2$ (superimposed lines) and the data from 
SN SDSS (top panel) and Union2.1 (bottom panel) SNe Ia compilations. Right panel: the relative differences of distance moduli in QSF+CDM and PSF+CDM with best-fit parameters determined from the same datasets 
(lines) are compared with the observational uncertainties for distance moduli presented in the left panels.}
\label{dl_all}
\end{figure}

\begin{figure}[tbp]
\centering
\includegraphics[width=.49\textwidth]{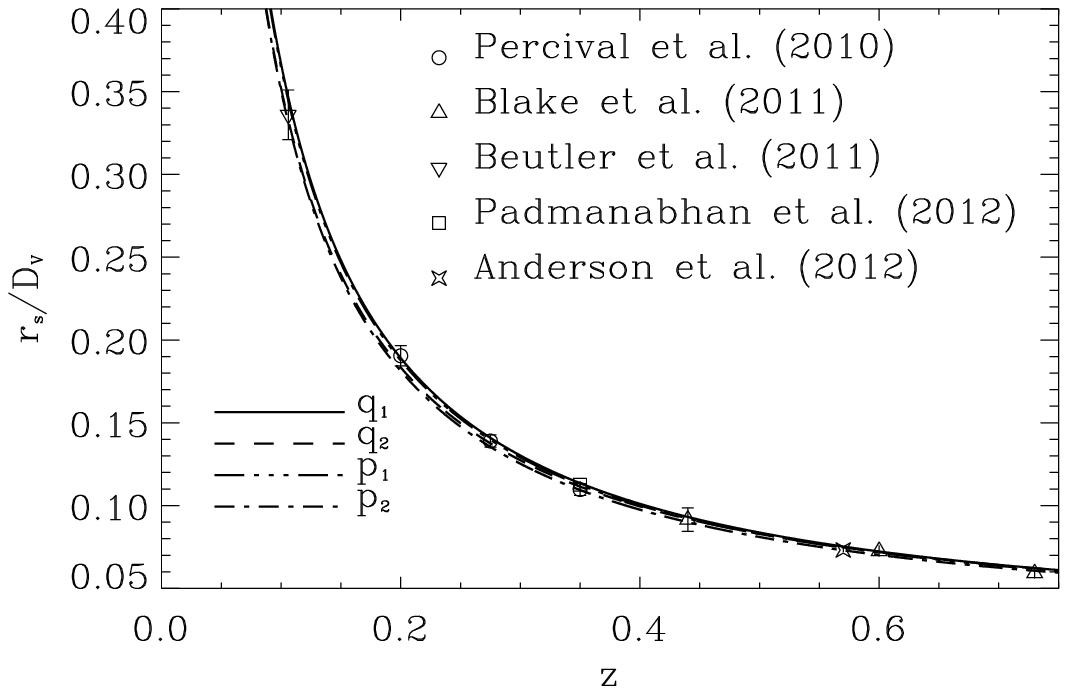}
\includegraphics[width=.49\textwidth]{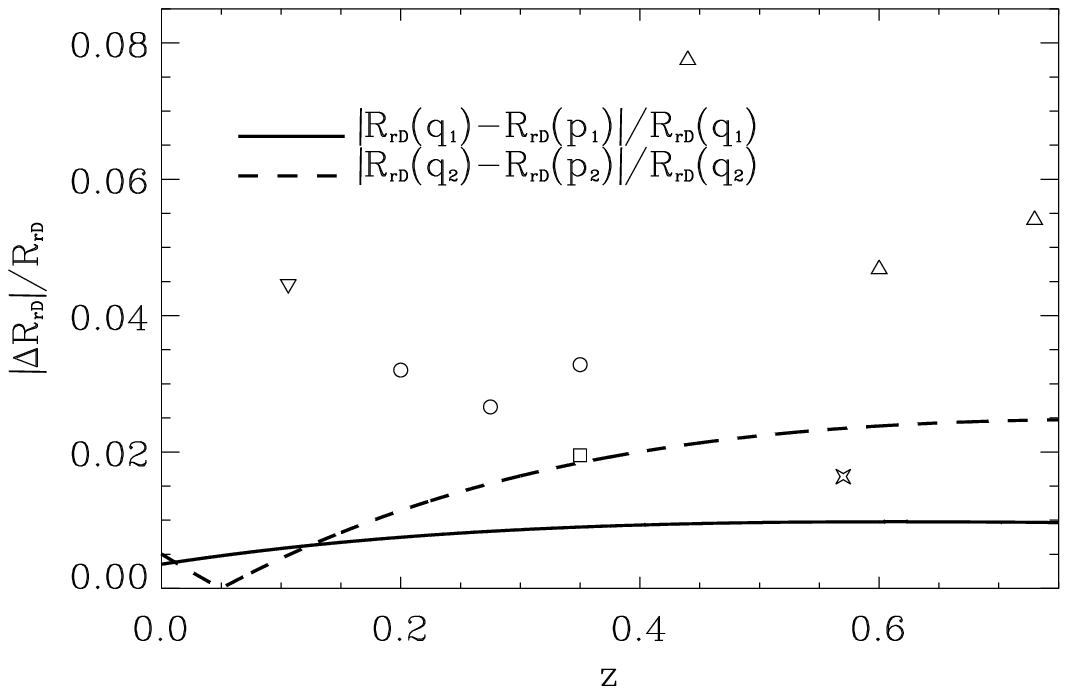}
\caption{Left panel: the relative BAO distance measure $R_{rD}\equiv r_s(z_{drag})/D_V(z)$ for the models with best-fit parameters 
$\mathbf{q}_1$, $\mathbf{q}_2$, $\mathbf{p}_1$ and $\mathbf{p}_2$ (lines) is compared with data from the SDSS DR7~\cite{Percival2010,Padmanabhan2012}, 
WiggleZ~\cite{wigglez}, 6dFGS~\cite{6dF} and SDSS DR9~\cite{Anderson2012} surveys (symbols). Right panel: the 
relative differences of the BAO distance measure $|\Delta R_{rD}|/ R_{rD}$ in the models with best-fit parameters $\textbf{q}_i$ and 
$\textbf{p}_i$. Symbols show the observational 1$\sigma$ errors of the data presented in the left panel.}
\label{rbao_all}
\end{figure}

\begin{figure}[tbp]
  \centering
  \includegraphics[width=.49\textwidth]{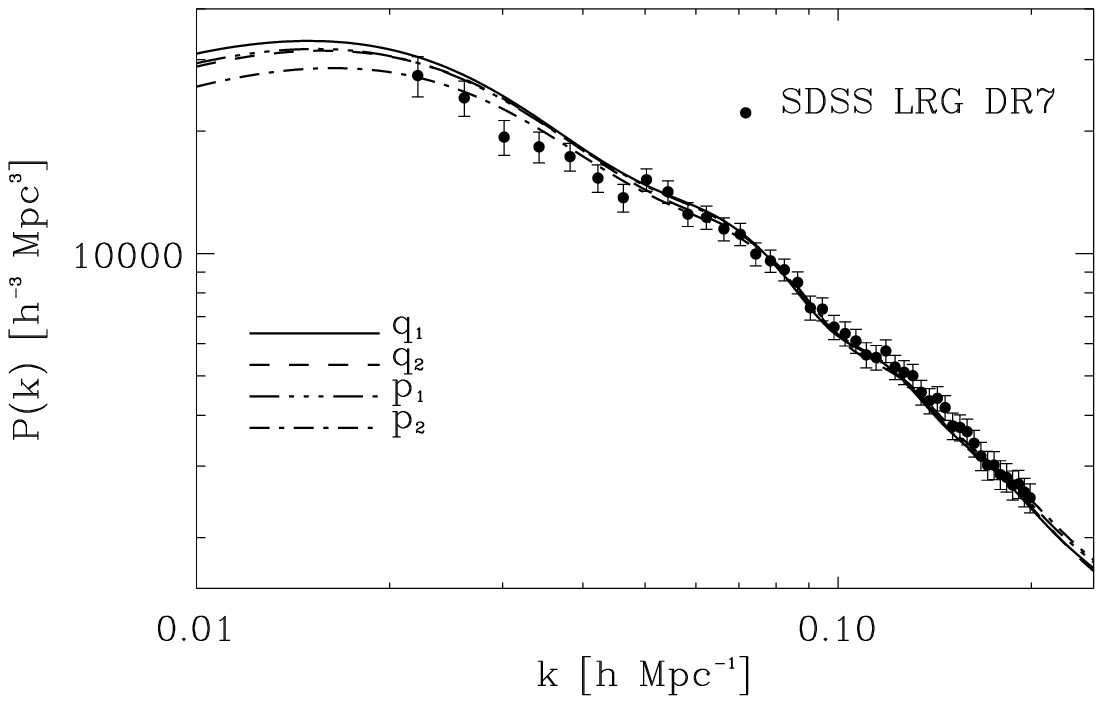}
  \includegraphics[width=.49\textwidth]{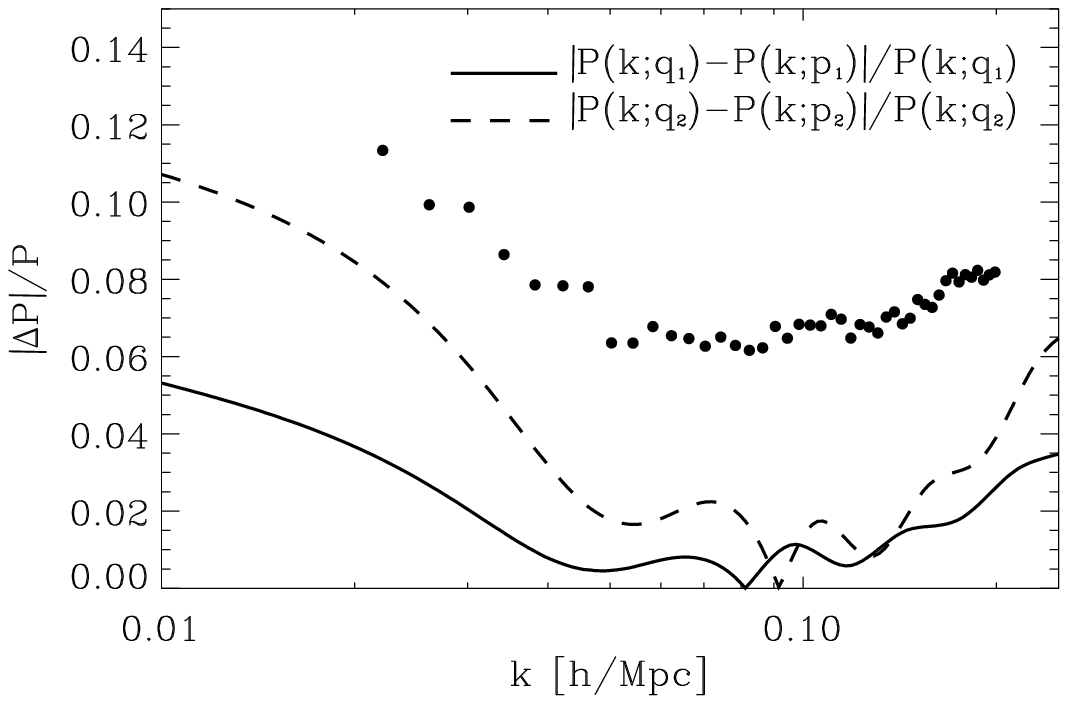}
  \caption{Left panel: the power spectrum of matter density perturbations in the cosmological models with best-fit parameters
    $\mathbf{q}_1$, $\mathbf{q}_2$, $\mathbf{p}_1$ and $\mathbf{p}_2$. Dots show the observational SDSS LRG DR7 power spectrum~\cite{Reid2010}. Right panel: the relative differences of matter density power spectra $|\Delta P(k)|/P(k)$ in
    the models with best-fit parameters $\textbf{q}_i$ and $\textbf{p}_i$. The dots show the observational uncertainties (1$\sigma$) of SDSS LRG DR7 data~\cite{Reid2010}.}
  \label{pk_all}
\end{figure}

\begin{figure}[tbp]
  \centering
  \includegraphics[width=0.95\textwidth]{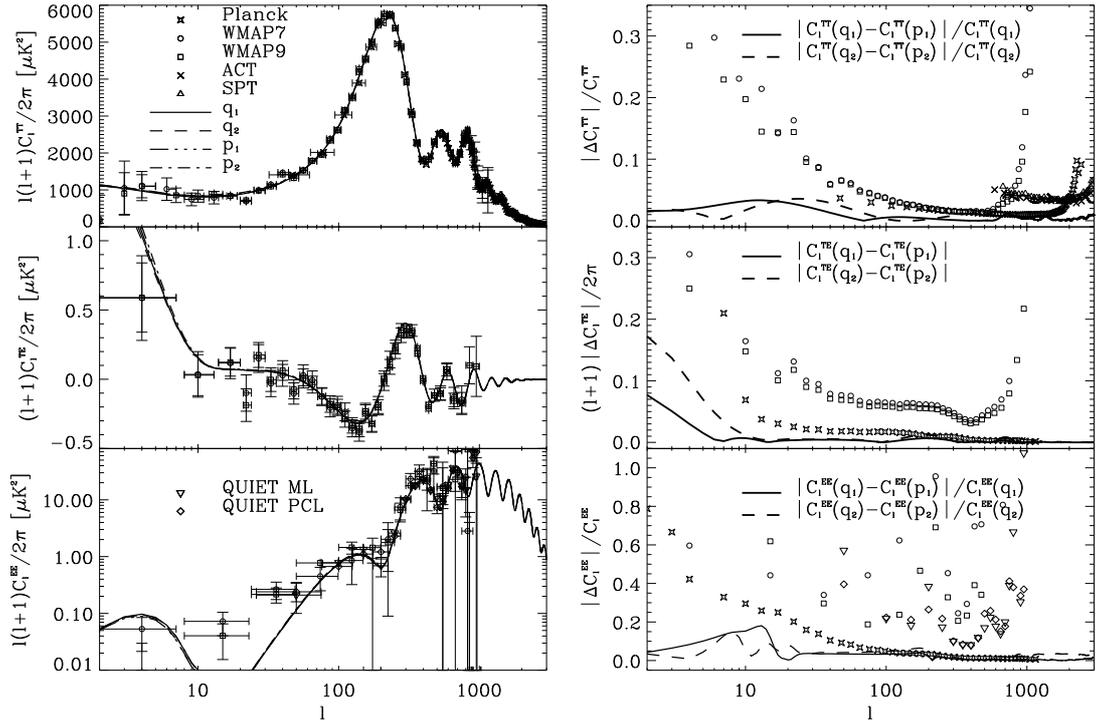}
  \caption{Left panel: the temperature (TT), temperature-polarization (TE) and polarization (EE) CMB angular power spectra,  $\ell(\ell+1)C^{TT}_{\ell}/2\pi$, $(\ell+1)C^{TE}_{\ell}/2\pi$ and $\ell(\ell+1)C^{EE}_{\ell}/2\pi$, for  the cosmological     models with best-fit parameters $\mathbf{q}_1$, $\mathbf{q}_2$, $\mathbf{p}_1$ and $\mathbf{p}_2$ (superimposed lines) are compared to  currently available data (symbols). Right panel: the relative differences of temperature $|\Delta C^{TT}_{\ell}|/C^{TT}_{\ell}$ and polarization $|\Delta C^{EE}_{\ell}|/C^{EE}_{\ell}$  power spectra and absolute differences for the temperature-polarization correlation, $(\ell+1)|\Delta C^{TE}_{\ell}|/2\pi$, in the models with best-fit parameters $\textbf{q}_i$ and $\textbf{p}_i$ (Table \ref{tab_bfp}). The symbols show 1$\sigma$ uncertainties of currently available and forecasted future data.}
  \label{cl_all}
\end{figure}

The QSF+CDM model with best-fit parameters $\textbf{q}_i$ and PSF+CDM model with best-fit parameters $\textbf{p}_i$ all provide a good fit to  
the data from SNe Ia distance moduli (left panels of Fig. \ref{dl_all}), BAOs (left panel of Fig. \ref{rbao_all}), matter power spectrum 
(left panel of Fig. \ref{pk_all}), CMB temperature fluctuations, polarization and cross-correlations polarization 
(left panels of Fig. \ref{cl_all}). Let us analyze the possibility to distinguish between QSF+CDM and PSF+CDM by each type of data.

The lines corresponding to different DE models in the left panel of Fig. \ref{dl_all} look perfectly superimposed. The maximal values of 
relative differences of SNe Ia distance moduli ($|\mu(\mathbf{q}_i)-\mu(\mathbf{p}_i)|/\mu(\mathbf{q}_i)$) 
for QSF+CDM and PSF+CDM models with best-fit parameters determined using the same fitters are less than $0.1\%$ (right panel of 
Fig.\ref{dl_all}). Comparing this to the observational uncertainties of SNe Ia distance moduli from SN SDSS and SN Union2.1 compilations 
(shown by dots, triangles and circles in the right panels) we conclude that these data are very far from distinguishing between the proposed 
types of scalar field dark energy. The same applies to the SNe Ia from SNLS3.

\begin{figure}[tbp]
  \centering
  \includegraphics[width=.49\textwidth]{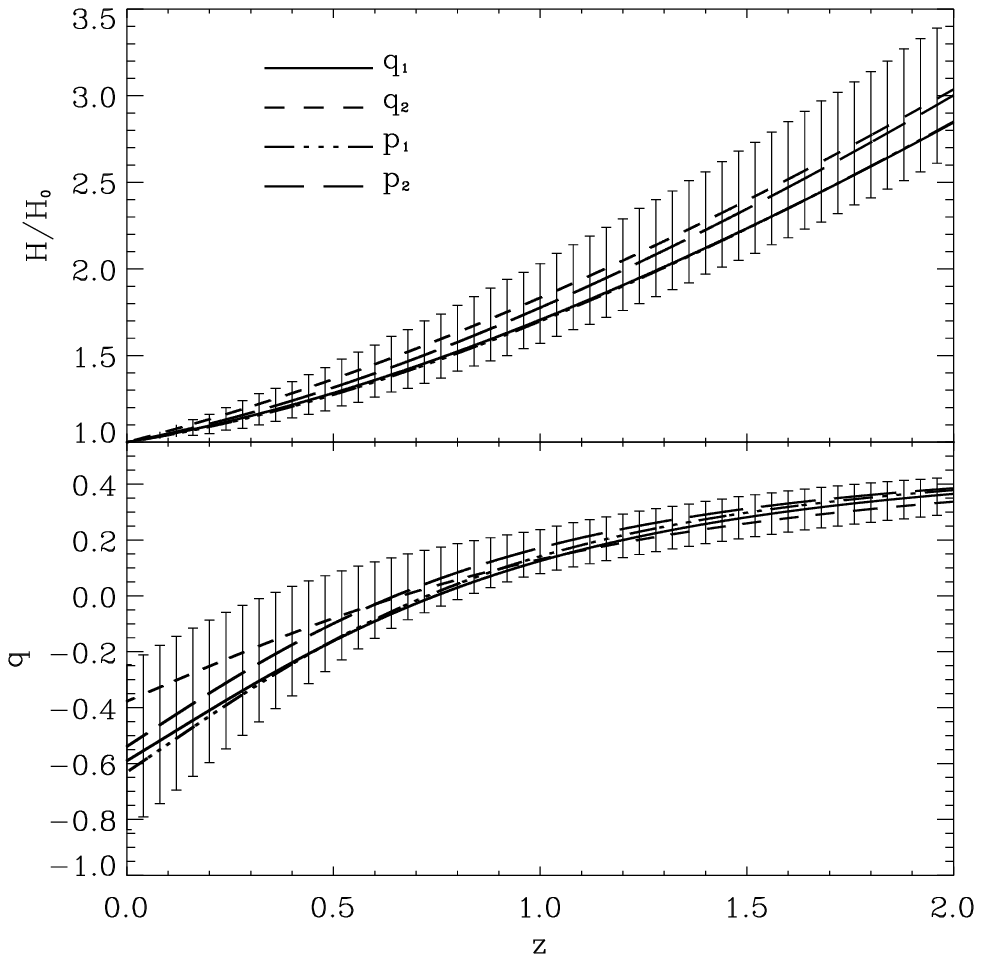}
  \includegraphics[width=.49\textwidth]{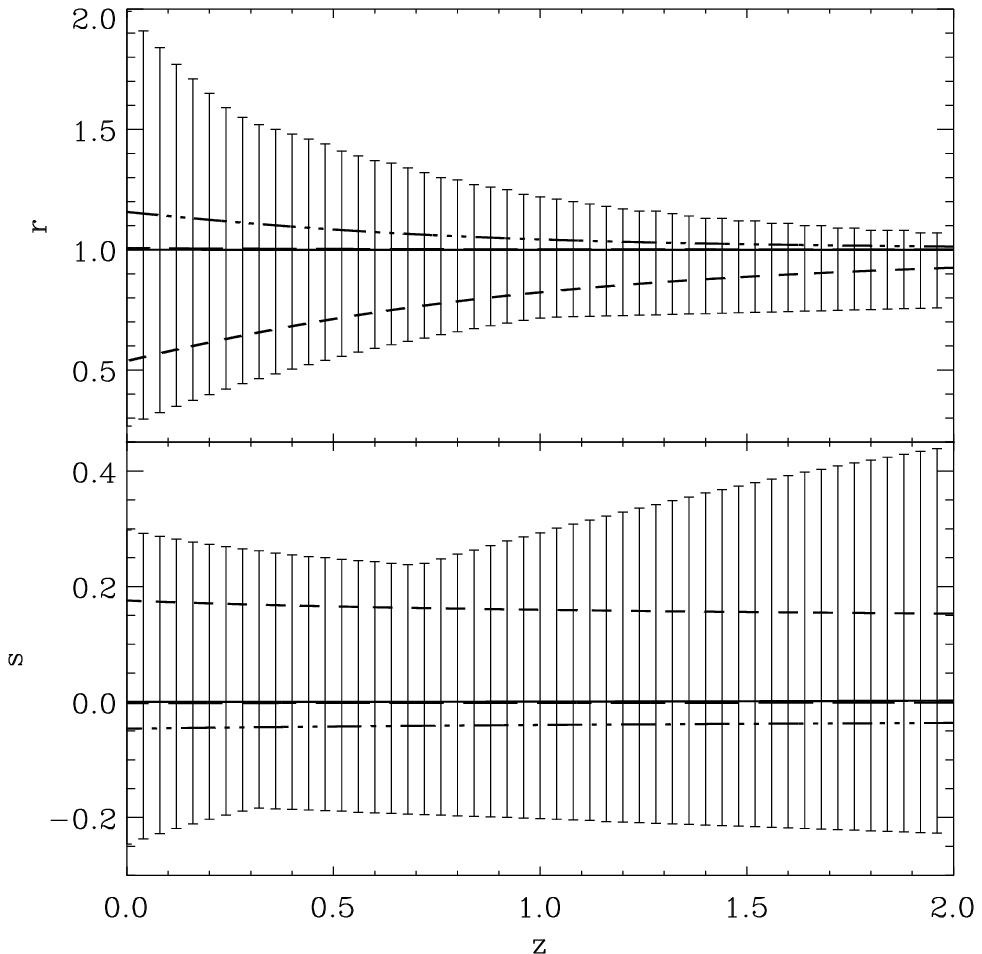}
  \caption{The redshift dependence of dimensionless parameters describing the expansion dynamics  of the Universe for the models with best-fit parameters $\mathbf{q}_1$, $\mathbf{q}_2$, $\mathbf{p}_1$ and $\mathbf{p}_2$. The rate of expansion $H/H_0$ (left panel, top), the deceleration parameter $q$ (left panel, bottom) and the statefinder parameters $r$ and $s$ (right panel) are shown. The dashed ranges are their observational uncertainties estimated by variation of the expressions (\ref{H_q}) for $H/H_0$, $q$, (\ref{rs_b}) for $r$ and $s$ over $\Omega_{de}$, $w_0$, $c_a^2$ for $2\sigma$ marginalized uncertainties of all models presented in Table \ref{tab_bfp}.}
  \label{qh_all}
\end{figure}

Fortunately, other characteristics of the expansion dynamics of the Universe, based on measurements of the first and second 
time derivatives of the Hubble parameter\footnote{$\mu(z)\equiv m-M=5\log{d_L}+25$ is the integral of $1/H$ over redshift, since $d_L(z)=(1+z)c\int_0^z dz'/H(z')$; $H(z)$ can be deduced from observations by differentiating of $\mu(z)$ with respect to $z$, since $H(z)=\left[\frac{d}{dz}(\frac{d_L}{z+1})\right]^{-1}$.} $H(z)$, are significantly more sensitive to the current value and time dependence of the EoS parameter. The redshift dependence of dimensionless parameters describing the  expansion dynamics of the Universe, such as the rate of expansion $H/H_0$, the deceleration parameter $q=-\dot{H}/(aH^2)-1$ and the statefinder parameters \cite{Sahni2003a}
\begin{equation}
r=\ddot{H}/(a^2H^3)+2\dot{H}/(aH^2)+1, \quad s\equiv (r-1)/3(q-1/2), \label{rs} 
\end{equation}
are shown in Fig. \ref{qh_all} for the models with best-fit parameters $\mathbf{q}_1$, $\mathbf{q}_2$, $\mathbf{p}_1$ and $\mathbf{p}_2$. In our dark energy models (\ref{w_rho}) the state finder parameters reduce to 
\begin{equation}
  r=1+4.5(1+w_{de})c_a^2\Omega_{de}(a), \quad\quad s=(1+w_{de})c^2_a/w_{de}, \label{rs_b} 
\end{equation}
where $\Omega_{de}(a)\equiv 8\pi G\rho_{de}(a)/3H^2$. The differences between $H/H_0$ for $\textbf{q}_i$ and $\textbf{p}_i$, as well as between $q(\textbf{q}_i)$ and $q(\textbf{p}_i)$, $s(\textbf{q}_i)$ and $s(\textbf{p}_i)$, $r(\textbf{q}_i)$ and $r(\textbf{p}_i)$, at different $z$ are significantly larger than those of $\mu(z)$ or $d_L(z)$. Unfortunately, current data on $d_L(z)$ from SN Ia measurements are too poor to determine  these quantities. The dashed ranges in 
Fig.~\ref{qh_all} show the dispersion obtained by the variation of expressions (\ref{H_q}) for $H/H_0$ and $q$, (\ref{rs_b}) for $r$ and $s$ over $\Omega_{de}$, $w_0$, $c_a^2$ within the $2\sigma$ uncertainties of all models presented in Table \ref{tab_bfp}. Maybe future high-precision measurements of distances to a significantly larger number of SN Ia or extremely well-calibrated $\gamma$-ray bursts will provide a possibility to distinguish the QSF and PSF models of DE. But clearly, present SNe Ia data is very far from this goal.

The relative BAO distance measure $R_{rD}(z)\equiv r_s(z_{drag})/D_V(z)$ (where $r_s(z_{drag})$ is the sound horizon at the epoch 
where photon drag stops due to recombination and $D_V(z)$ is the BAO dilation scale) extracted from SDSS DR7 (see \cite{Percival2010} 
and improved analysis \cite{Padmanabhan2012}), 
the WiggleZ survey~\cite{wigglez}, 6dFGS~\cite{6dF} and the SDSS DR9~\cite{Anderson2012} galaxy redshift surveys is matched well by both models, 
QSF+CDM and PSF+CDM  for both sets of best-fit parameters $\mathbf{q}_1$, $\mathbf{q}_2$, $\mathbf{p}_1$ and $\mathbf{p}_2$ 
(see Fig.~\ref{rbao_all}, left panel). The relative differences of $R_{rD}(z)$ for models with QSF and PSF 
are up to $\sim1-2.5\%$ for $0.1\le z\le0.8$, while observational errors are about $2-3.5\%$ for SDSS DR7 BAO data, $\sim5-8\%$ for 
WiggleZ one, $\sim4\%$ for 6dF one and somewhat smaller than $2\%$ for SDSS DR9 one (right panel of Fig. \ref{rbao_all}). 
Note that the relative difference of $\mathbf{q}_2$ and $\mathbf{p}_2$ is larger than the relative uncertainty of SDSS DR9 BAO data and 
comparable to the relative uncertainty of improved SDSS DR7 BAO data. This means that an improvement in the determination of 
$R_{rD}(z)$ in future large scale structure surveys like BigBOSS~\cite{bigboss} or Euclid~\cite{euclid} can distinguish between these 
types of dark energy.

\begin{figure}[tbp]
\centering
\includegraphics[width=.99\textwidth]{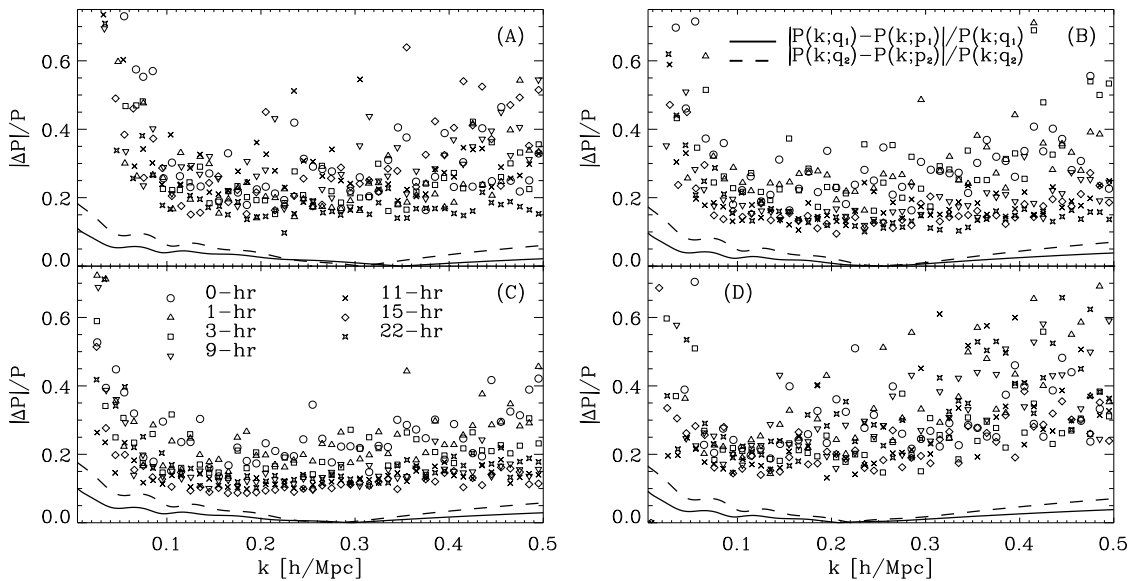}
\caption{The relative differences of the matter density power spectra $|\Delta P(k)|/P(k)$ in comparison with the WiggleZ power spectrum 
uncertainties for galaxy samples at different redshift bins (A: $0.1<z<0.3$, B: $0.3<z<0.5$, C: $0.5<z<0.7$, D: $0.7<z<0.9$) and from different 
regions in the sky (symbols).}
  \label{dpk_wigglez}
\end{figure}

The power spectrum of matter density perturbations extracted from luminous red galaxies sample from SDSS DR7 catalogue by Reid et al. (2010) 
\cite{Reid2010} has not been used for the determination of best-fit parameters of QSF+CDM and PSF+CDM, however the computed power spectra 
match it well too. The experimental errors of its determination are still too large ($8-12\%$) to distinguish between different scalar field 
models of dark energy. The same applies to the power spectra of galaxies from the WiggleZ survey \cite{Parkinson2012}, for which the r
elative uncertainties are $\gtrsim 10-20\%$ (Fig.~\ref{dpk_wigglez}). In addition, at small scales ($k\ge0.1$ hMpc$^{-1}$) there are uncertainties in the computation of the power 
spectrum associated with the nonlinear evolution of perturbations. The resolution of this problem would require combined N-body + scalar field 
simulations.


Above we have discussed the importance of CMB data  for the determination of cosmological parameters and, in particular, for dark energy parameters. The key cosmological data for the last years have been the WMAP all-sky maps, which contain information about the primordial fluctuation amplitude and spectral index, the positions and amplitudes of the acoustic peaks,  as well as the amplitude of large scale matter density perturbations at late time imprinted on the integrated Sachs-Wolfe effect. In Fig.~ \ref{cl_all} (top left panel) the power spectra of temperature fluctuations extracted from the 7-year \cite{WMAP7a,WMAP7b,WMAP7c} and 9-year \cite{wmap9a,wmap9b} WMAP all-sky measurements are shown. Its accuracy is best (minimal errors $\sim 1.5-4\%$) in the range of the first and second acoustic peaks ($\ell\sim 200-600$). This allows an accurate determination of the main cosmological parameters. The best accuracy of dark energy parameters is achieved when CMB data are used together with SNe Ia data and  
relative BAO 
distance measures or directly the matter power spectrum. The power spectra $\ell(\ell+1)C^{TT}_{\ell}/2\pi$ for the QSF+CDM and PSF+CDM models with best-fit parameters $\mathbf{q}_1$, $\mathbf{q}_2$, $\mathbf{p}_1$ and $\mathbf{p}_2$ provide good fits to the WMAP7 and WMAP9 data, as well as to other recent CMB experiments, ACT \cite{act} and SPT \cite{spt}, which are sensitive to higher values of $\ell$. 
The relative difference between the spectra in models with parameters $\mathbf{q}_2$ and $\mathbf{p}_2$, shown in the top right panel 
of Fig. \ref{cl_all}, is comparable to the relative uncertainties of WMAP7 and especially WMAP9 data at $\ell\sim 300-800$ and to the relative 
uncertainties of ACT and SPT data at higher $\ell$.

\begin{figure}[tbp]
  \centering
  \includegraphics[width=0.75\textwidth]{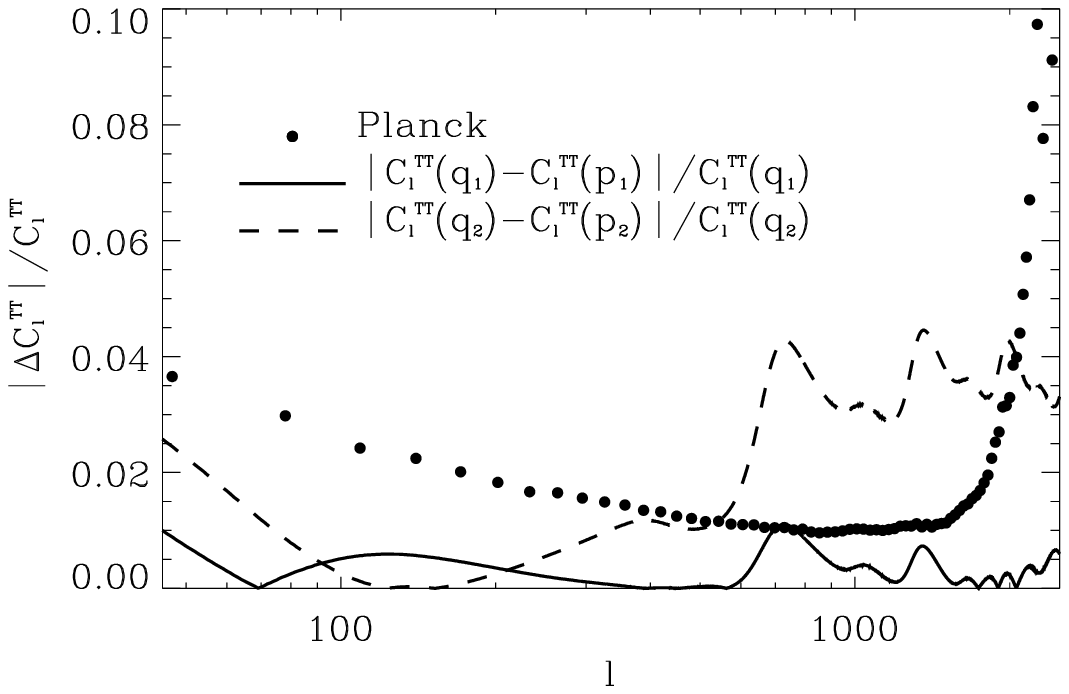}
  \caption{The relative differences of temperature anisotropy $|\Delta C^{TT}_{\ell}|/C^{TT}_{\ell}$ power spectra in the models with best-fit parameters $\textbf{q}_i$ and $\textbf{p}_i$. The symbols show 1$\sigma$ uncertainties of Planck data.}
  \label{diff_planck}
\end{figure}

We also present the uncertainties of the Planck CMB temperature fluctuations power spectrum for $\ell\sim50-2500$ \cite{Planck}. 
At high $\ell$ the Planck uncertainties are smaller than the difference between models with parameters $\mathbf{q}_2$ and 
$\mathbf{p}_2$ (at $\ell\sim 500-2000$) and comparable to the difference between models with parameters $\mathbf{q}_1$ and $\mathbf{p}_1$ (at $\ell\sim 700-800$) (see Fig.~\ref{diff_planck}). This is in agreement with our 
previous conclusion \cite{Novosyadlyj2011,Sergijenko2011} that the accuracy of the power spectra expected from Planck will significantly narrow 
the allowed range of parameter  values for scalar field models of dark energy.

Additional constraints on cosmological parameters are obtained when CMB polarization data are included. To illustrate the agreement between theory and observations, the temperature-polarization $(\ell+1)C^{TE}_{\ell}/2\pi$ and polarization $\ell(\ell+1)C^{EE}_{\ell}/2\pi$ CMB power spectra for the cosmological models with best-fit parameters  $\mathbf{q}_1$, $\mathbf{q}_2$, $\mathbf{p}_1$ and $\mathbf{p}_2$ are presented in the middle and bottom left panels of Fig. \ref{cl_all}. All lines are superimposed for $\ell>20$ with sub-percent accuracy for TE and a few percents for EE, while the errors of the WMAP7 and WMAP9 power spectra at these $\ell$'s are larger than $\sim 3-6\%$ for TE and  $\sim 20-40\%$ for EE, as it is shown in the middle and bottom right panels of Fig.~\ref{cl_all}. In the bottom right panel of Fig. \ref{cl_all} we present also the relative uncertainties from the QUIET experiment \cite{quiet} (for both maximal likelihood and pseudo-$C_{\ell}$s). The minimal errors of the QUIET EE power 
spectra at high $\ell$ are at least $\sim 10-20\%$. We also present the forecasted uncertainties of the Planck data (based on the sensitivities for 7 frequency channels which can measure the polarization \cite{bluebook}, $f_{sky}$ is taken to be 0.65). They are comparable to the differences between the models at high $\ell$. All this means that the polarization data  of the CMB are not sufficient to distinguish between quintessence and phantom fields at the present level of accuracy of the observational data, but they will possibly become sufficient in the near future.

\section{Conclusion}

Among the used observational data only the CMB temperature fluctuations power spectrum from Planck and the baryon acoustic oscillations from SDSS DR9 can marginally distinguish between quintessence (QSF {+} CDM) and phantom (PSF {+} CDM) cosmologies at a statistically significant level. In the framework of each model a set of best-fit parameters exists. The best-fit model matches well each type of data and all together with similar goodness of fit. The data on CMB temperature fluctuations from WMAP9, ACT and SPT and the expected Planck polarization data as well as the improved data on BAO from SDSS DR7 look promising for the purpose of distinguishing between these models of dark energy. In the future, increasing the accuracy of CMB polarization measurements jointly with high precision matter density data will probably allow us to to distinguish between PSF and QSF dark energy, and may establish the dynamical properties of dark energy and, maybe, its nature. On the other hand, SNe Ia luminosity distance 
measurements do not seem very promising.

\appendix
\section{Observational constraints on scalar field models with \texorpdfstring{$w_0>-1$, $c_a^2<-1$}{w0ca2}}\label{app}

Another subclass of scalar field models of dark energy with barotropic equation of state without peculiarities in the past are fields with $w_0>-1$, $c_a^2<-1$. In the past the EoS parameter of such fields evolved from $-1$ at early epoch to $w_0$ today. The properties of these models will be studied in detail in a separate paper, here we only present observational constraints on them from the datasets 1 and 2, determined by the MCMC technique in the same way as in \cite{Novosyadlyj2012}.

\begin{figure}[tbp]
\centerline{\includegraphics[width=\textwidth]{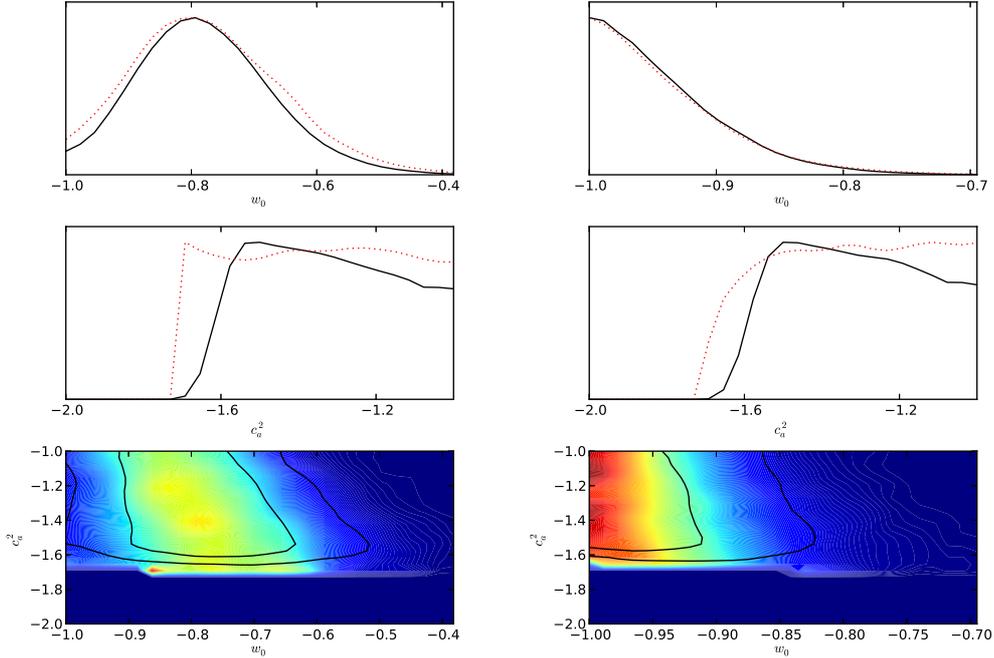}} 
\caption{One-dimensional marginalized posteriors (solid lines) and mean likelihoods (dotted lines) for $w_0$ (top panels) and $c_a^2$ (middle panels) in models with $w_0>-1$, $c_a^2<-1$. Left: WMAP7 {+} HST {+} BBN {+} BAO {+} SN SDSS MLCS2k2. 
Rright: WMAP7 {+} HST {+} BBN {+} BAO {+} SN SDSS SALT2. Bottom: the corresponding two-dimensional mean likelihood distributions in the plane $c_a^2-w_0$. Solid lines show the $1\sigma$ and $2\sigma$ confidence contours.}
\label{postlike_qp}
\end{figure}

In Fig. \ref{postlike_qp} we see that while for $w_0$ the posteriors and mean the likelihoods are close and have the shape of Gaussian and half-Gaussian, for $c_a^2$ the shapes of the likelihoods differ significantly from each other and from Gaussian. This means that also for such fields the range of acceptable values of $c_a^2$ is generally unconstrained (it is constrained only in the case $\mathbf{q}_2$). The ranges of acceptable values of all other parameters of the models studied here including $w_0$ are well constrained. In Table \ref{tab_qp} we present the best-fit values and $2\sigma$ marginalized confidence ranges of cosmological parameters in models with scalar fields with $w_0>-1$, $c_a^2<-1$. The value of $\chi^2$ for this model in case of dataset 1 is larger than the corresponding value for both quintessence and phantom dark energy. On the other hand, for the dataset 2 the value of $\chi^2$ is smaller than for the quintessence and phantom models. Thus the scalar field with $w_0>-1$, 
$c_a^2<-1$ is preferred by the dataset 2, however, again the difference is not statistically significant.

\begin{table}[tbp]
  \centering
  \caption{The best-fit values, mean values and 2$\sigma$ marginalized confidence ranges for parameters of cosmological models with scalar fields with $w_0>-1$, $c_a^2<-1$ determined by the MCMC technique using two observational datasets: WMAP7 {+} HST {+} BBN {+} BAO {+} SN SDSS SALT2 and WMAP7 {+} HST {+} BBN {+} BAO {+} SN SDSS MLCS2k2. The rescaled energy density of the component $X$ is denoted by $\omega_X \equiv \Omega_Xh^2$.}
  \medskip
 \begin{tabular}{|c|c|c|c|c|}
 \hline
&\multicolumn{2}{c|}{}&\multicolumn{2}{c|}{}\\
Parameter&\multicolumn{2}{c|}{SALT2}&\multicolumn{2}{c|}{MLCS2k2}\\
&\multicolumn{2}{c|}{}&\multicolumn{2}{c|}{}\\
\cline{2-5}&&&&\\
&best-fit&2$\sigma$ c.l.&best-fit&2$\sigma$ c.l.\\
&&&&\\
 \hline
&&&&\\
$\Omega_{de}$& 0.727& 0.725$_{- 0.030}^{+ 0.027}$& 0.700& 0.699$_{- 0.034}^{+ 0.031}$\\&&&&\\
$w_0$&-0.987&-0.939$_{- 0.061}^{+ 0.098}$&-0.849&-0.777$_{- 0.159}^{+ 0.184}$\\&&&&\\
$c_a^2$&-1.481&-1.313$_{- 0.252}^{+ 0.313}$&-1.073&-1.329$_{- 0.259}^{+ 0.329}$\\&&&&\\
$10\omega_b$& 0.224& 0.226$_{- 0.010}^{+ 0.011}$& 0.226& 0.227$_{- 0.010}^{+ 0.011}$\\&&&&\\
$\omega_{cdm}$& 0.112& 0.111$_{- 0.007}^{+ 0.007}$& 0.111& 0.111$_{- 0.008}^{+ 0.009}$\\&&&&\\
$h$& 0.701& 0.697$_{- 0.026}^{+ 0.025}$& 0.668& 0.665$_{- 0.027}^{+ 0.030}$\\&&&&\\
$n_s$& 0.968& 0.971$_{- 0.024}^{+ 0.025}$& 0.972& 0.973$_{- 0.025}^{+ 0.026}$\\&&&&\\
$\log(10^{10}A_s)$& 3.084& 3.082$_{- 0.066}^{+ 0.071}$& 3.086& 3.084$_{- 0.066}^{+ 0.070}$\\&&&&\\
$\tau_{rei}$& 0.087& 0.089$_{- 0.023}^{+ 0.025}$& 0.090& 0.090$_{- 0.023}^{+ 0.025}$\\&&&&\\
 \hline
&\multicolumn{2}{c|}{}&\multicolumn{2}{c|}{}\\
$-\log L$&\multicolumn{2}{c|}{ 3865.05}&\multicolumn{2}{c|}{ 3857.09}\\&\multicolumn{2}{c|}{}&\multicolumn{2}{c|}{}\\
 \hline
 \end{tabular}
\label{tab_qp}
\end{table}

Comparing the relative uncertainties of SNe Ia distance moduli, BAO relative distance measures, matter density perturbations and CMB power spectra with differences between the corresponding quantities in such models and quintessence or phantom, we find that scalar fields with $w_0>-1$, $c_a^2<-1$ can only be distinguished from quintessence or phantom by Planck and SDSS DR9 BAO data. The probability of distinguishing them from phantom fields is slightly better than from quintessence.

\acknowledgments

This work was supported by the project of Ministry of Education and Science of Ukraine (state registration number 0113U003059), 
research program ``Scientific cosmic research'' of the National Academy of Sciences of Ukraine (state registration number 0113U002301) and the 
SCOPES project No. IZ73Z0128040 of Swiss National Science Foundation. Authors also acknowledge the use of CAMB and CosmoMC packages.

\end{document}